  \documentstyle[aas2pp4,epsf]{article}

\newcommand{\Ell}{E_\parallel}      
\newcommand{\rhoGJ}{\rho_{{\rm GJ}}}  
\newcommand{\Bc}{B_{\rm cnt}}      
\newcommand{\sgP}{\sigma_{\rm p}}  
\newcommand{\rlc}{\varpi_{\rm LC}} 
\newcommand{\Rc}{R_{\rm C}}        
\newcommand{\Ex}{\epsilon_{\rm x}} 
\newcommand{\Eg}{\epsilon_\gamma}  
\newcommand{\inc}{\alpha_{\rm i}}  
\newcommand{\figA}{\mbox fig.~1}  
\newcommand{\figB}{\mbox fig.~3}  

\lefthead{Hirotani and Shibata}
\righthead{}

\begin{document}

\title{Electrodynamic Structure of an Outer--Gap Accelerator:
       Implausibility of Super Goldreich-Julian Current}
\author{Kouichi Hirotani}
 \affil{NASA/Goddard Space Flight Center, 
        Greenbelt, MD~20771\\
       hirotani@hotaka.mtk.nao.ac.jp}
\and
\author{Shinpei Shibata}
\affil{Department of Physics, Yamagata University,
       Yamagata 990-8560, Japan\\
       shibata@sci.kj.yamagata-u.ac.jp}

\begin{abstract}
We investigate a stationary pair production cascade 
in the outer magnetosphere of a spinning neutron star.
The charge depletion due to global flows of charged particles,
causes a large electric field along the magnetic field lines.
Migratory electrons and/or positrons are accelerated by this field
to radiate curvature gamma-rays, 
some of which collide with the X-rays to materialize as pairs in the gap.
The replenished charges partially screen the electric field, 
which is self-consistently solved, 
together with the distribution functions of particles and gamma-rays.
By solving the Vlasov equations describing this pair production cascade,
we demonstrate the existence of a stationary gap in the outer 
magnetosphere of PSR B1055-52 
for a wide range of current densities flowing in the accelerator:
From sub  to super Goldreich-Julian values. 
However, we find that the expected GeV spectrum becomes very soft
if the current density exceeds the Goldreich-Julian value.
We also demonstrate that the GeV spectrum softens
with decreasing magnetic inclination 
and with increasing distance to this pulsar.
We thus conclude that a sub-Goldreich-Julian current, 
a large magnetic inclination, and a small distance (500 pc, say)
are plausible to account for EGRET observations.
Furthermore, it is found that the TeV flux 
due to inverse Compton scatterings of
infrared photons whose spectrum is inferred from the Rayleigh-Jeans
side of the soft blackbody spectrum
is much less than the observational upper limit.
\end{abstract}

\keywords{gamma-rays: observations -- gamma-rays: theory -- 
          -- magnetic fields 
          -- pulsars: individual (B1055--52)
          -- X-rays: galaxies}


\section{Introduction}
\label{sec:intro}

The EGRET experiment on the Compton Gamma Ray Observatory
has detected pulsed signals from seven rotation-powered pulsars
(for Crab, Nolan et al. 1993, Fierro et al. 1998;
 for Vela, Kanbach et al. 1994, Fierro et al. 1998;
 for Geminga, Mayer-Hasselwander et al. 1994, Fierro et al. 1998; 
 for PSR B1706--44, Thompson et al. 1996;
 for PSR B1046--58, Kaspi at al. 2000;
 for PSR B1055--52, Thompson et al. 1999;
 for PSR B1951+32, Ramanamurthy et al. 1995).
The modulation of the $\gamma$-ray light curves at GeV energies 
testifies to the production of $\gamma$-ray radiation in the pulsar 
magnetospheres either at the polar cap 
(Harding, Tademaru, \& Esposito 1978; Daugherty \& Harding 1982, 1996;
 Michel 1991;
 Sturner, Dermer, \& Michel 1995;
 Shibata, Miyazaki, \& Takahara 1998;
 Harding \& Muslimov 1998; 
 Zhang \& Harding 2000),
or at the vacuum gaps in the outer magnetosphere
(e.g., Cheng, Ho, \& Ruderman 1986a,b, hereafter CHRa,b).
Both of these pictures have had some success in reproducing global 
properties of the observed emission.
However, there is an important difference between these two models:
An polar gap accelerator releases very little angular momenta,
while outer gap one could radiate them efficiently.
More specifically, the total angular momentum loss rate must equal
the energy loss rate divided by the angular velocity of the star,
implying an average location of energy loss on the light cylinder
(Cohen \& Treves 1972; Holloway 1977; Shibata 1995),
of which distance from the rotation axis is given by
\begin{equation}
  \rlc = \frac{c}{\Omega},
\end{equation}
where $\Omega$ denotes the angular frequency of the neutron star, 
and $c$ the speed of light.

On these grounds, the purpose here is to explore a little further
into a model of an outer gap, which is located near the light cylinder.
If the outer magnetosphere is filled with a plasma so that the space 
charge density is equal to the Goldreich-Julian density, 
$\rho_{\rm GJ} \equiv -\Omega B_{\rm z} / (2\pi c)$, 
then the field-aligned electric field vanishes,
where $B_{\rm z}$ is the component of the magnetic field 
along the rotational axis.
However, the depletion of charge in the Goldreich and Julian model
in a region where it could not be resupplied,
may cause a vacuum region to develop.
Holloway (1973) pointed out the possibility that a region which lacks
plasma is formed around the surface on which $\rho_{\rm GJ}$ 
changes its sign.
Subsequently,
CHRa,b developed a version of an outer magnetospheric $\gamma$-ray 
emission zone in which acceleration in the Holloway gaps above 
the null surface, where $B_{\rm z}$ vanishes,
brought particles to large Lorentz factors ($\sim 10^{7.5}$).

After CHRa,b, many outer-gap models have been proposed
(Chiang \& Romani 1992, 1994; Romani \& Yadigaroglu 1995;
 Romani 1996; Zhang \& Cheng 1997; Cheng, Ruderman \& Zhang 2000).
Their basic idea is as follows:
A deviation of the charge density from $\rhoGJ$ results in an electric 
field along the magnetic field, ${\bf B}$.
If this electric field becomes strong enough to accelerate $e^\pm$ pairs 
to ultrarelativistic energies, they could radiate $\gamma$-rays tangential 
to the curved ${\bf B}$ field lines there.
The curvature $\gamma$-rays may be converted into $e^\pm$ pairs via 
two photon collisions.
In order to keep a steady current flow and the charge density $\rhoGJ$ 
in the regions outside the gap, the gap will grow until it is large 
enough and the electric field is strong enough to maintain a sufficient 
supply of charges to the rest of the open field line region.
If the gap ends in a region where $\rhoGJ \ne 0$ and 
${\bf E}\cdot{\bf B} \ne 0$, charges from the surrounding region may flow 
in through the end (or the boundary).
If both boundaries are located on the null surface, 
pairs produced in the gap will replace the charge deficiency 
inside the gap, and finally the gap will 
be filled up.
However, if a vacuum gap extends to the light cylinder, 
the charged particles created in the gap should escape 
from the magnetosphere, so the gap would not be quenched.
Hence, stable outer gaps (if they exist) are those from the null surface 
to the light cylinder along the last-open field lines.
In an outer gap, the inner boundary lies near the intersection of the 
null surface where $\rhoGJ$ vanishes 
and the boundary of the closed field lines 
of the star on which the magnetospheric current does not exits.

Recently, solving the Vlasov equations that describe a stationary 
pair production cascade, Hirotani and Shibata (2001a,b, Papers VII and VIII) 
elucidated important characteristics of the particle acceleration zones 
in the outer magnetosphere.
They demonstrated that a stationary gap does not extend between 
the null surface and the light cylinder.
Rather, their width ($W$) along the field lines is adjusted so that 
a single particle emits copious $\gamma$-rays one of which materialize 
as a pair in the gap on average.
The resultant $W$ becomes about $5\%$ of $\rlc$ in the case of 
the Crab pulsar.
The produced pairs in the gap do not completely cancel the acceleration field,
because the particles do not accumulate at the gap boundaries.
Outside of the gap, the particles flows away from the gap along the
magnetic field lines as a part of the global current flow pattern.
In another word, the remained, small-amplitude electric field
along the magnetic field lines outside of the gap
prevents the gap from quenching.
Moreover, they demonstrated that the gap position is not fixed 
as considered in previous outer gap models. 
Their position shifts as the magnetospheric current flowing the gap changes.
That is, if there is no particle injection across either of the boundaries,
the gap is located around the null surface, 
as demonstrated in their earlier papers
(Hirotani \& Shibata 1999a,b,c, Papers~I, II, III; 
 Hirotani 2000a,b,c, Papers~IV,V,VI;
 see also Beskin et al. 1992 and Hirotani \& Okamoto 1998 for
 a pair-production avalanche in a black-hole magnetosphere).
However, the gap position shifts towards the light cylinder 
(or the star surface) if the injected particle flux across the inner 
(or the outer) boundary approaches the Goldreich-Julian value.
In other words, the accelerator can be located at any place in the 
magnetosphere; their position is 
primarily constrained by the magnetospheric current.

In Papers~VII and VIII, they examined the outer gaps formed in the 
magnetospheres of young or millisecond pulsars, assuming a 
sub-Goldreich-Julian particle injection across the boundaries.
In this paper, we apply the same method to a middle-aged pulsar B1055-52 
and demonstrate that a stationary gap is maintained 
even if the injected particle flux exceeds the Goldreich-Julian value, 
and that the possibility of such a large particle flux injection is 
ruled out if we compare the predicted $\gamma$-ray spectrum with observations.

In the next section, we formulate the basic equations describing
the stationary pair production cascade in the outer magnetosphere.
In \S~3, we apply the method to a middle-aged pulsar, B~1055-52
and reveal that large magnetic inclination and a sub-Goldreich-Julian current 
are plausible for the outer-magnetospheric accelerator for this pulsar.
In the final section, we compare the results with previous works. 

\section{Basic Equations}
\label{sec:basic}

We first consider the continuity equations of particles 
in \S~\ref{sec:cont},
the $\gamma$-ray Boltzmann equations
in \S~\ref{sec:Boltz_gamma},
and the Poisson equation for the electrostatic potential 
in \S~\ref{sec:Poisson}.

\subsection{Particle Continuity Equations}
\label{sec:cont}

Under the mono-energetic approximation, 
we simply assume that the electrostatic and the 
curvature-radiation-reaction forces cancel each other
in the Boltzmann equations of particles.
Then the spatial number density of positrons ($e^+$'s) and 
electrons ($e^-$'s), $N_+(s)$ and $N_-(s)$, 
obey the following continuity equations:
\begin{equation}
  \frac{\partial N_\pm}{\partial t} 
  + \frac{\partial}{\partial \vec{x}} \left( \vec{v}_\pm N_\pm \right)
  = Q(\vec{x}),
  \label{eq:cont-eq-0}
\end{equation}
where
\begin{equation}
  Q(\vec{x}) \equiv
  \frac{1}{c} \int_{0}^\infty d\epsilon_\gamma \, 
    [ \eta_{\rm p+} G_+   +\eta_{\rm p-} G_- ].
  \label{eq:def_Q}
\end{equation}
Here, 
$G_+(\vec{x},\epsilon_\gamma)$ and $G_-(\vec{x},\epsilon_\gamma)$ 
refer to the distribution functions of
outwardly and inwardly propagating $\gamma$-ray photons,
respectively, having energy $m_{\rm e}c^2 \epsilon_\gamma$;
$\vec{v}_+$ (or $\vec{v}_-$) designates the velocity of the
center of mass of $e^+$'s (or $e^-$'s).
The pair production rate for an outwardly propagating
(or inwardly propagating) $\gamma$-ray photon to materialize 
as a pair per unit time is expressed by 
$\eta_{{\rm p}+}$ (or $\eta_{{\rm p}-}$).
For charge definiteness, we consider that 
a positive electric field arises in the gap.
In this case, 
positrons (or electrons) are migrating outwardly (or inwardly).

Since gyration will be canceled out in $\vec{v}_\pm$,
we obtain
\begin{equation}
  \vec{v}_\pm= \pm c\cos\Phi\frac{\vec{B}}{B}
               +   r\Omega\sin\theta \vec{e}_\phi,
  \label{eq:def_v}
\end{equation}
where
\begin{equation}
  \vec{e}_\phi
  \equiv \frac{1}{r\sin\theta} \frac{\partial}{\partial \phi}
  \label{eq:def_ephi}
\end{equation}
is the azimuthal unit vector,
$\Phi \equiv \sin^{-1}(r\Omega\sin\theta/c)$ 
is the projection angle of the three-dimensional particle velocity
onto the poloidal plane,
and $B \equiv \vert \vec{B} \vert$.
Imposing the stationary condition $\partial_t + \Omega\partial_\phi=0$,
we obtain from equations~(\ref{eq:cont-eq-0}) and (\ref{eq:def_Q})
\begin{equation}
  \pm \vec\nabla \cdot \left( c\cos\Phi \frac{\vec{B}}{B} N_\pm \right) 
  = \int_{0}^\infty d\epsilon_\gamma \, 
    [ \eta_{\rm p+} G_+   +\eta_{\rm p-} G_- ].
  \label{eq:cont-eq-1}
\end{equation}
In this paper, we estimate $\Phi$ at the gap center such that
\begin{equation}
  \Phi= \sin^{-1} \frac{r_{\rm cnt} \Omega \sin\theta_{\rm cnt}}{c},
  \label{eq:def_Phi}
\end{equation}
where $r_{\rm cnt}$ is the distance of the gap center
from the star center,
and $\theta_{\rm cnt}$ is its colatitude angle.
Utilizing $\vec{\nabla} \cdot \vec{B}=0$,
and assuming that the toroidal bending is negligible in the sense that
$\vert B_\phi \vert \ll B$,
we obtain
\begin{equation}
  \pm B \frac{d}{ds}\left( \frac{N_\pm}{B} \right)
  = \frac{1}{c \cos\Phi} \int_{0}^\infty d\epsilon_\gamma \, 
    [ \eta_{\rm p+} G_+   +\eta_{\rm p-} G_- ].
  \label{eq:cont-eq}
\end{equation}

The pair production redistribution functions $\eta_{{\rm p}\pm}$ 
can be defined as
\begin{equation}
  \eta_{{\rm p}\pm}(\Eg)
  = (1-\mu_{\rm c}) c
     \int_{\epsilon_{\rm th}}^\infty d\epsilon_{\rm x}
     \frac{dN_{\rm x}}{d\Ex} 
     \sgP(\Eg,\Ex,\mu_{\rm c}),
  \label{eq:def_etap_0}
\end{equation}
where $\sgP$ is the pair-production cross section and
$\cos^{-1}\mu_{\rm c}$ refers to the collision angle between 
the $\gamma$-rays and the X-rays with energy $m_{\rm e}c^2 \Ex$.
X-ray number density between  dimensionless energies
$\Ex$ and $\Ex+d\Ex$,
is integrated over $\Ex$ from
the threshold energy
$\epsilon_{\rm th} \equiv 2(1-\mu_{\rm c})^{-1} \Eg^{-1}$
to infinity.

To evaluate $\mu_{\rm c}$ between the surface X-rays and the $\gamma$-rays,
we must consider $\gamma$-ray's toroidal momenta due to the aberration
of light.
It should be noted that a $\gamma$-ray photon propagates 
in the instantaneous direction of the particle motion 
at the time of the emission.
Therefore, evaluating the $\gamma$-ray toroidal velocity 
at the gap center,
we obtain
\begin{equation}
  \mu_{\rm c}= \cos\Phi \cos\theta_{\rm pol},
  \label{eq:def_coll_surface}
\end{equation}
where $\theta_{\rm pol}$ is the collision angle between 
the surface X-rays and the curvature $\gamma$-rays,
that is, the angle between the two vectors
($r_{\rm cnt}$,$\theta_{\rm cnt}$) and 
($B^r$,$B^\theta$) at the gap center.
Therefore, the collision approaches head-on (or tail-on) 
for inwardly (or outwardly) propagating $\gamma$-rays, 
as the gap shifts towards the star.

\subsection{Boltzmann Equations for Gamma-rays}
\label{sec:Boltz_gamma}

Unlike the charged particles,
$\gamma$-rays do not propagate along the magnetic field line at each point,
because they preserve the directional information where they were emitted.
However, to avoid complications, we simply assume 
that the outwardly (or inwardly) propagating $\gamma$-rays
dilate (or constrict) at the same rate with the magnetic field;
this assumption gives a good estimate 
when $W \ll \rlc$ holds.
Under this assumption, we obtain (Paper~VI)
\begin{eqnarray}   
  \pm B \frac{\partial}{\partial s} \left( \frac{G_\pm}{B}\right)
     = - \frac{\eta_{{\rm p}\pm}}{c\cos\Phi} G_\pm
       + \frac{\eta_{\rm c}     }{c\cos\Phi} N_\pm,
  \label{eq:Boltz_gam}
\end{eqnarray}   
where (e.g., Rybicki, Lightman 1979)
\begin{equation}
  \eta_{\rm c} \equiv \frac{\sqrt{3}e^2 \Gamma}{h \Rc}
               \frac1{\epsilon_\gamma} 
	       F \left( \frac{\epsilon_\gamma}{\epsilon_{\rm c}} \right) ,
  \label{eq:def-etaC}
\end{equation}
\begin{equation}
  \epsilon_{\rm c} \equiv \frac1{m_{\rm e}c^2} \frac3{4\pi}
                          \frac{hc \Gamma^3}{\Rc} ,
  \label{eq:def_Ec}
\end{equation}
\begin{equation}
  F(u) \equiv u \int_u^\infty K_{\frac53} (t) dt ;
\end{equation}
$\Rc$ is the curvature radius of the magnetic field lines
and $K_{5/3}$ is the modified Bessel function of $5/3$ order.
The effect of the broad spectrum of curvature $\gamma$-rays
is represented by the factor $F(\epsilon_\gamma/\epsilon_{\rm c})$
in equation (\ref{eq:def-etaC}).
$\Gamma$ refers to the particle Lorentz factor.

\subsection{Poisson Equation}
\label{sec:Poisson}

The real charge density $e(N_+ -N_-)$ appears in the Poisson equation
for the non-corotational potential $\Psi$. 
Neglecting relativistic effects,
we can write down the Poisson equation as follows:
\begin{equation}
 -\nabla^2 \Psi 
    = 4\pi \left[ e(N_+ -N_-) +\frac{\Omega B_z}{2\pi c} \right],
  \label{eq:Poisson_0}
\end{equation}
where $e$ designates the magnitude of the charge on an electron.

\subsection{One-dimensional Analysis}
\label{sec:reduction}

As described at the end of \S~3 in Paper~VII, 
it is convenient to introduce the 
typical Debey scale length $c/\omega_{\rm p}$, 
\begin{equation}
  \omega_{\rm p} = \sqrt{ \frac{4\pi e^2}{m_{\rm e}}
	                  \frac{\Omega \Bc}{2\pi ce} },
  \label{eq:def-omegap}
\end{equation}
where $\Bc$ represents the magnetic field strength at the gap center.
The dimensionless coordinate variable then becomes
\begin{equation}
  \xi \equiv (\omega_{\rm p}/c) s.
  \label{eq:def-xi}
\end{equation}
By using such dimensionless quantities, 
and by assuming that the transfield thickness, $D_\perp$,
of the gap is greater than $W$,
we can rewrite the Poisson equation~(\ref{eq:Poisson_0}) into
the following one-dimensional form
(e.g., \S~2 in Michel 1974):
\begin{equation}
  E_\parallel = -\frac{d\psi}{d\xi},
  \label{eq:basic-1}
\end{equation}
\begin{equation}
  \frac{dE_\parallel}{d\xi}
  = - \frac{\psi}{\Delta_\perp^2}
    + \frac{B(\xi)}{\Bc} \left[ n_+(\xi) - n_-(\xi) \right]
    + \frac{B_z(\xi)}{\Bc}
  \label{eq:basic-2}
\end{equation}
where $ \psi(\xi) \equiv e\Psi(s)/(m_{\rm e}c^2)$
and $\Delta_\perp \equiv (\omega_{\rm p}/c) D_\perp$;
the particle densities are normalized by the Goldreich-Julian value as
\begin{equation}
  n_\pm(\xi) \equiv 
    \frac{2\pi ce}{\Omega} \frac{N_\pm}{B}.
  \label{eq:def-n}
\end{equation}
We evaluate $B_z/B$ at each point along the last-open field line,
by using the Newtonian dipole field.

Let us introduce the following dimensionless $\gamma$-ray 
densities in the dimensionless energy interval
between $\beta_{i-1}$ and $\beta_i$:
\begin{equation}
  g_\pm^i(\xi) \equiv 
    \frac{2\pi ce}{\Omega \Bc}
    \int_{\beta_{i-1}}^{\beta_i} d\epsilon_\gamma G_\pm(s,\epsilon_\gamma).
  \label{eq:def-g}
\end{equation}
In this paper, we set $\beta_0=10^2$,
which corresponds to the lowest $\gamma$-ray energy, $51.1$ MeV.
We divide the $\gamma$-ray spectra into $11$ energy bins 
and put
$\beta_1= 10^{2.5}$, 
$\beta_2= 10^3$, 
$\beta_3= 10^{3.25}$, 
$\beta_4= 10^{3.5}$, 
$\beta_5= 10^{3.75}$, 
$\beta_6= 10^{4}$, 
$\beta_7= 10^{4.25}$, 
$\beta_8= 10^{4.5}$,
$\beta_9= 10^{4.75}$,
$\beta_{10}= 10^{5}$,
$\beta_{11}= 10^{5.25}$.
We can now rewrite the continuity quation~(\ref{eq:cont-eq}) 
of particles into 
\begin{equation}
  \frac{dn_\pm}{d\xi} = 
    \pm \frac{\Bc}{B\cos\Phi}
        \sum_{i=1}^{9} [ \eta_{\rm p+}{}^i g_+^i(\xi)
                        +\eta_{\rm p-}{}^i g_-^i(\xi)],
  \label{eq:basic-3}
\end{equation}
where the magnetic field strength, $B$, is evaluated at each $\xi$.
The dimensionless redistribution functions
$\eta_{{\rm p}\pm}^i$ 
are evaluated at the central energy in each bin as
\begin{equation}
  \eta_{{\rm p}\pm}^i \equiv
  \frac{1}{\omega_{\rm p}}
  \eta_{{\rm p}\pm}\left(\frac{\beta_{\rm i-1}+\beta_{\rm i}}{2}\right).
  \label{eq:def_etap_1}
\end{equation} 

A combination of equations (\ref{eq:basic-3}) 
gives the current conservation law,
\begin{equation}
  j_{\rm tot} \equiv n_+(\xi) + n_-(\xi) = {\rm constant \ for \ } \xi.
  \label{eq:consv}
\end{equation}
If $j_{\rm tot}=1$ holds, 
the current density per unit magnetic flux tube equals the
Goldreich-Julian value, $\Omega/(2\pi)$.

The Boltzmann equations~(\ref{eq:Boltz_gam}) for the $\gamma$-rays
are integrated over $\Eg$ between
dimensionless energies $\beta_{\rm i-1}$ and $\beta_{\rm i}$ 
to become
\begin{eqnarray}   
  \frac{d}{d\xi} g_\pm^i
     = \frac{d}{d\xi}\left( \ln B \right) g_\pm^i
       \mp \frac{\eta_{{\rm p}\pm}{}^i}{\cos\Phi} g_\pm^i
       \pm \frac{\eta_{\rm c}^i B(\xi)}{\Bc\cos\Phi} n_\pm,
  \label{eq:basic-5}
\end{eqnarray}   
where $i=1,2,\cdot\cdot\cdot,m$ ($m=9$) and 
\begin{eqnarray}
  \eta_{\rm c}^i 
  &\equiv& \frac{\sqrt{3}e^2\Gamma}{\omega_{\rm p}hR_{\rm c}}
           \int_{\beta_{i-1} / \epsilon_{\rm c}}
               ^{\beta_i     / \epsilon_{\rm c}}
            du \int_u^\infty K_{\frac53}(t)dt
  \label{eq:etaCi}
\end{eqnarray}
is dimensionless.

Equating the electric force $e \vert d\Psi / dx \vert$ and the
radiation reaction force,
we obtain the saturated Lorentz factor at each point as 
\begin{equation}
  \Gamma_{\rm sat} 
   = \left( \frac{3 R_{\rm c}{}^2}{2e} 
		    \left| \frac{d\Psi}{ds} \right|
                  + 1 
     \right)^{1/4};
  \label{eq:saturated}
\end{equation}
we compute the curvature radius $R_{\rm c}$ 
at each point for a Newtonian dipole magnetic field.
In the case of middle-aged pulsars such as B1055-52,
their less dense X-ray fields lead to a large pair-production
mean free path for a $\gamma$-ray photon to materialize as a pair.
To ensure that the Lorentz factor do not exceed the maximum attainable
limit, $\psi(\xi_2)$, we compute $\Gamma$ with
\begin{equation}
  \frac{1}{\Gamma}
  = \sqrt{ \frac{1}{\Gamma_{\rm sat}{}^2}
          +\frac{1}{\psi^2(\xi_2)}
         }.
  \label{eq:terminal}
\end{equation}

\subsection{Boundary Conditions}
\label{sec:BD}

We now consider the boundary conditions 
to solve the Vlasov equations
(\ref{eq:basic-1}), (\ref{eq:basic-2}), (\ref{eq:basic-3}), 
and (\ref{eq:basic-5}).
At the {\it inner} (starward) boundary
($\xi= \xi_1$), we impose (Paper~VI)
\begin{equation}
  E_\parallel(\xi_1)=0,
  \label{eq:BD-1}
\end{equation}
\begin{equation}
  \psi(\xi_1) = 0,
  \label{eq:BD-2}
\end{equation}
\begin{equation}
  g_+^i(\xi_1)=0  \quad (i=1,2,\cdot\cdot\cdot,9).
  \label{eq:BD-3}
\end{equation}
It is noteworthy that condition (\ref{eq:BD-1}) is consistent with
the stability condition at the plasma-vacuum interface 
if the electrically supported magnetospheric plasma
is completely-charge-separated, 
i.e., if the plasma cloud at $\xi < \xi_1$ is composed of 
electrons alone (Krause-Polstorff \& Michel 1985a,b; Smith et al. 2001).
We assume that the Goldreich-Julian plasma gap boundary 
is stable with $\Ell=0$ on the boundary, $\xi=\xi_1$.

Since positrons may flow into the gap at $\xi=\xi_1$
as a part of the global current pattern in the magnetosphere,
we denote the positronic current per unit flux tube at $\xi=\xi_1$ as
\begin{equation}
  n_+(\xi_1)= j_1,
  \label{eq:BD-4}
\end{equation}
which yields (eq.~[\ref{eq:consv}])
\begin{equation}
  n_-(\xi_1)= j_{\rm tot}-j_1.
  \label{eq:BD-5}
\end{equation}

At the {\it outer} boundary ($\xi=\xi_2$), we impose
\begin{equation}
  E_\parallel(\xi_2)=0,
  \label{eq:BD-6}
\end{equation}
\begin{equation}
  g_-^i(\xi_2)=0 \quad (i=1,2,\cdot\cdot\cdot,9),
  \label{eq:BD-7}
\end{equation}
\begin{equation}
  n_-(\xi_2)= j_2.
  \label{eq:BD-8}
\end{equation}

The current density created in the gap per unit flux tube
can be expressed as
\begin{equation}
  j_{\rm gap}= j_{\rm tot} -j_1 -j_2.
  \label{eq:Jgap}
\end{equation}
We adopt $j_{\rm gap}$, $j_1$, and $j_2$
as the free parameters.

We have totally $2m+6$ boundary conditions 
(\ref{eq:BD-1})--(\ref{eq:BD-8})
for $2m+4$ unknown functions
$\Psi$, $E_\parallel$,
$n_\pm$, 
$g_\pm^i$ ($i=1, 2, \cdot\cdot\cdot, m$),
where $m=11$.
Thus two extra boundary conditions must be compensated 
by making the positions of the boundaries $\xi_1$ and $\xi_2$ be free.
The two free boundaries appear because $E_\parallel=0$ is imposed at 
{\it both} the boundaries and because $j_{\rm gap}$ is externally imposed.
In other words, the gap boundaries ($\xi_1$ and $\xi_2$) shift,
if $j_1$ and/or $j_2$ varies.

It is worth mentioning that the gap width, $W$, is 
related with the X-ray field density as follows
(for details, see Papers~V, VII, and VIII):
\begin{equation}
  W = \frac{\lambda_{\rm p}}{N_\gamma}
      \frac{j_{\rm gap}}{j_{\rm tot}},
  \label{eq:closure}
\end{equation}
where $\lambda_{\rm p}$ refers to the pair production optical depth,
and $N_\gamma$ the expectation value of the $\gamma$-ray photons
emitted by a single particle while it runs the gap.
Equation~(\ref{eq:closure})
is automatically satisfied by the stationary Vlasov equations. 
The relation is useful when we interpret the variation of $W$.
For example, for middle-aged pulsars,
the X-ray field is much less dense compared with the 
magnetospheric X-ray field of young pulsars.
We can therefore expect that the gap is more extended 
(i.e., $W$ is greater) for middle-aged pulsars
than the young pulsars,
because the former's larger $\lambda_{\rm p}$ leads to a larger $W$,
compared with the latter.
We examine such features more accurately in the next section,
by analyzing the Vlasov equations numerically
under the boundary conditions (\ref{eq:BD-1})--(\ref{eq:BD-8}).

\section{Application to PSR B~1055-52}
\label{sec:app_B1055}

\subsection{X-ray and Infrared Field}
\label{sec:XIR_B1055}

Combining ROSAT and ASCA data, Greiveldinger et al. (1996)
reported that the X-ray spectrum consists of two components:
a soft blackbody with $kT_{\rm s}=68$~eV and 
$A_{\rm s}= 0.78 A_\ast (d/0.5)^2$
and a hard blackbody with $kT_{\rm h}=320$~eV and 
$A_{\rm h}= 2.5 \times 10^{-5} A_\ast (d/0.5)^2$,
where $d$ refers to the distance in kpc. 
Here, $A_{\rm s}$ and $A_{\rm h}$ indicate the observed emission
erea of soft and hard blackbody components, respectively;
$A_\ast$ expresses the area of the whole neutron star surface.

Mignani, Caraveo, and Bignami (1997) claimed that 
the blue/UV flux is consistent with an extrapolation of the 
soft, thermal part of the ROSAT X-ray spectrum.
We thus compute the infrared flux from the Rayleigh-Jeans side
of the soft X-ray spectrum with $kT_{\rm s}=68$~eV and 
$A_{\rm s}= 0.78 A_\ast (d/0.5)^2$. 
This IR photon field is, in fact, so weak
that the IC drag acting on a particle is negligibly small
compared with the curvature drag,
as the TeV/GeV flux ratio ($\ll 1$) indicates in the following 
subsections.
Therefore, the Vlasov equations can be solved without the IR field
(i.e., the pair-production rate can be computed only from the X-ray field).
Once $\Ell(\xi)$, $n_\pm(\xi)$, and $g_\pm{}^i(\xi)$ are solved,
we can passively compute the TeV spectrum.

\subsection{Dependence on Injected Currents}
\label{sec:res_B1055}

Let us now substitute the X-ray field into
equation~(\ref{eq:def_etap_0}) and 
solve the Vlasov equations by the method described in \S~\ref{sec:basic}.
The explicit GeV and TeV spectra are computed by the method described 
in \S~4 in Paper VII.

\subsubsection{Equal Current Injection}
\label{sec:j1=j2_B1055}

\begin{figure} 
\centerline{ \epsfxsize=8.5cm \epsfbox[200 20 500 250]
              {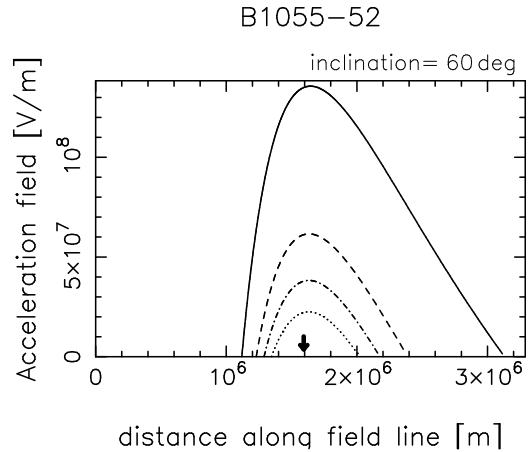} } 
\caption{\label{fig:Ell_1055_60_01_j1=j2} 
Distribution of the acceleration field, $-d\Psi/ds$, for PSR~B1055-52 
for $\inc= 60^\circ$.
The solid, dashed, dash-dotted, and dotted curves correspond to
the cases of ($j_{\rm gap}$,$j_1$,$j_2$)$=$(0.01,0,0),
(0.01,0.005,0.005), (0.01,0.05,0.05), and (0.01,2,2), respectively.
The downarrow indicates the position of the null surface,
where $B_z$ vanishes.
        }
\end{figure}

We first consider the case when $j_1=j_2$.
In figure~\ref{fig:Ell_1055_60_01_j1=j2},
we present the acceleration field,
\begin{equation}
  -\frac{d\Psi}{ds}
  = \frac{\omega_{\rm p}}{c}
    \frac{m_{\rm e}c^2}{e} \Ell,
  \label{eq:acc_field}
\end{equation}
for the four cases:
$j_1=j_2=0$, $0.005$, $0.05$, and $2.0$,
which are indicated by the solid, dashed, dash-dotted, and dotted 
curves, respectively.
In the third case for instance,
the positronic and electronic currents 
equally flow into the gap
across the inner and outer boundaries at the rate 
$0.05\Omega /2\pi$ per unit magnetic flux tube
(i.e., $5\%$ of the typical Goldreich-Julian value).
For all the four cases, we set $\inc=60^\circ$.
Since the solution forms a \lq brim' 
to disappear (fig.~2 in Hirotani \& Okamoto 1998)
if $j_{\rm gap}$ exceeds typically several percent,
we adopt $j_{\rm gap}= 0.01$ as the representative value.
The abscissa designates the distance along the last-open field line
and covers the range from the neutron star surface ($s=0$)
to the position where the disance equals 
$s= 0.35 \rlc= 3.29 \times 10^6$~m.

It follows from figure~\ref{fig:Ell_1055_60_01_j1=j2} that
the gap is located around the null surface (where $B_z$ vanishes),
which is indicated by the downarrow.
For $\inc=60^\circ$, $\Omega=31.9 \,\mbox{rad s}^{-1}$,
the null surface is located at $s= 1.59 \times 10^6$~m from the 
star surface along the last-open field line.
The conclusion that the gap is located around the null surface
for $j_1=j_2$, 
is consistent with what was obtained analytically in \S~2.4 in Paper~VII.
We can also understand from figure~\ref{fig:Ell_1055_60_01_j1=j2} 
that $W$, and hence $-d\Psi/ds$
decreases with increasing $j_1=j_2$.
This is because not only the produced particles in the gap 
(i.e., $j_{\rm gap}$) 
but also the injected ones (i. e., $j_1+j_2$)
contribute to the $\gamma$-ray emission.
Because of the efficient $\gamma$-ray emission due to the injected 
particles, 
a stationary pair-production cascade is maintained with a small 
pair-production optical depth $W/\lambda_{\rm p}$,
and hence a small $W$.
It is noteworthy that $W$ does not linearly decrease with 
$j_{\rm gap}/j_{\rm tot}$,
by virtue of the \lq negative feedback effect' 
due to the $N_\gamma^{-1}$ factor;
that is, the decrease in $N_\gamma$ due to the reduced $W$ 
partially cancel the original decrease in $W$.
It may be noteworthy that $N_\gamma \propto W\Gamma \propto W^{3/2}$ 
holds (for details, see eqs.~[8], [11], and [31] in Paper~V,
along with fig.~6 in Hirotani \& Okamoto 1998).

\begin{figure} 
\centerline{ \epsfxsize=8.5cm \epsfbox[200 20 500 250]
             {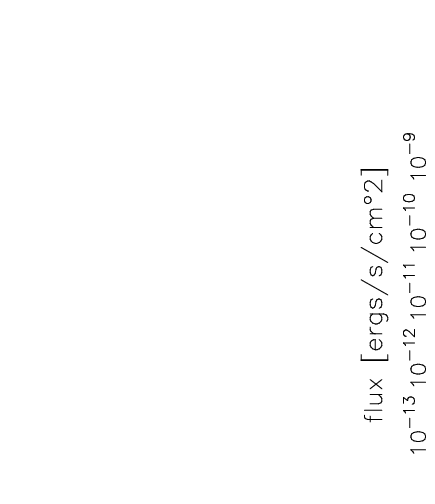} } 
\caption{\label{fig:Spc_1055_60_01_j1=j2}
Expected pulsed $\gamma$-ray spectra from PSR~B1055-52
for $\inc= 60^\circ$ and $j_{\rm gap}=0.01$.
The four curves correspond to the same cases as presented in \figA.
In each case, the larger flux for 
either the inwardly or the outwardly propagating $\gamma$-rays
is depicted.
        }
\end{figure}

We present the $\gamma$-ray spectrum
in figure~\ref{fig:Spc_1055_60_01_j1=j2};
the curves correspond to the same cases as those 
in figure~\ref{fig:Ell_1055_60_01_j1=j2}. 
In each case, outwardly propagating $\gamma$-ray flux is depicted, 
because it is greater than the inwardly propagating one. 
The observed fluxes are indicated by open circles and squares.
It follows that the spectrum becomes soft
as the injected current densities, $j_1=j_2$, increase.
This can be easily understood because $-d\Psi/ds$
decreases with increasing $j_1=j_2$ 
(fig.~\ref{fig:Ell_1055_60_01_j1=j2}).
Even though solutions exist for super Goldreich-Julian currents
(e.g., the dotted line in fig.~\ref{fig:Spc_1055_60_01_j1=j2}), 
the predicted spectra become too soft to match the observations.
On these grounds, we can conclude that 
{\it a sub-Goldreich-Julian current 
density is preferable for us to account for the observed spectrum}.

\subsubsection{Current Injection across Outer Boundary}
\label{sec:in_B1055}

\begin{figure} 
\centerline{ \epsfxsize=8.5cm \epsfbox[200 20 500 250]
              {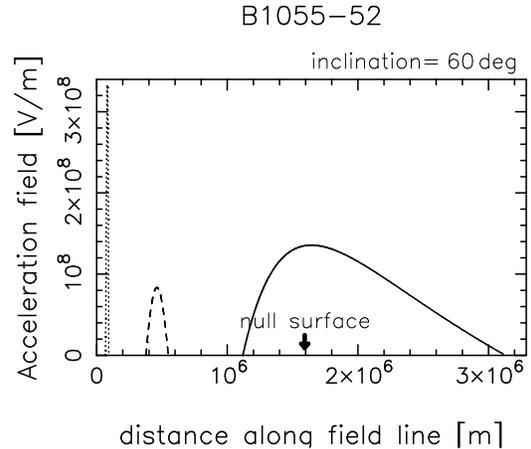} } 
\caption{\label{fig:Ell_1055_60_01_00_j2} 
Distribution of $-d\Psi/ds$ for PSR~B1055-52 for $\inc= 60^\circ$. 
The solid, dashed, and dotted curves correspond to
the cases of ($j_{\rm gap}$,$j_1$,$j_2$)$=$(0.01,0,0),
(0.01,0,0.25), and (0.01,0,0.391), respectively.
        }
\end{figure}

We next consider the case when the gap is shifted towards the star surface
by virture of the particle injection across the outer boundary.
In figure~\ref{fig:Ell_1055_60_01_00_j2},
we present $-d\Psi/ds$ for vanishing $j_1$;
the solid, dashed, and dotted curves
represent the cases of
($j_1$,$j_2$)= (0,0), (0,0.25), and (0,0.391), respectively.
For all the three cases, we set
$\inc=60^\circ$ and $j_{\rm gap}=0.01$.
The abscissa covers the range from $s=0$ to $s= 0.35 \rlc$.
There is no solution above $j_2>0.391$, 
because $-d\Psi/ds$ distribution
forms a brim at the inner boundary at this critical value.

It follows from the figure that the gap position shifts inwards
as $j_2$ increases.
For example, the gap center position shifts from
$0.225\rlc$ for $j_2=0$ (solid curve) to  
$0.049\rlc$ and $0.0084\rlc$ for $j_2= 0.25$ (dashed)
and $j_2=0.391$ (dotted), respectively,
from the star surface along the last-open field line.
In addition, $W$ decreases significantly as the gap shifts inwards.
This is because both the factors $j_{\rm gap}/j_{\rm tot}$ and
$\lambda_{\rm p}$ decreases in equation~(\ref{eq:closure}).
For details, see \S~6.1 in Paper~VII. 

Let us now consider the Lorentz factor.
As the maximum value in the gap, 
we obtain $3.0 \times 10^7$ for $j_2=0$ (solid curve), whereas
$8.7 \times 10^6$ for $j_2=0.391$ (dotted one).
That is, in the latter case,
$\Gamma$ reduces significantly even though the maximum of $-d\Psi/ds$
is greater than the former.
This is because the particles' motion becomes unsaturated 
due to the reduced $W$.
Therefore, the mono-enegetic approximation is no longer valid for
the dotted curve in figure~\ref{fig:Ell_1055_60_01_00_j2}.

\begin{figure} 
\centerline{ \epsfxsize=8.5cm \epsfbox[200 20 500 250]
             {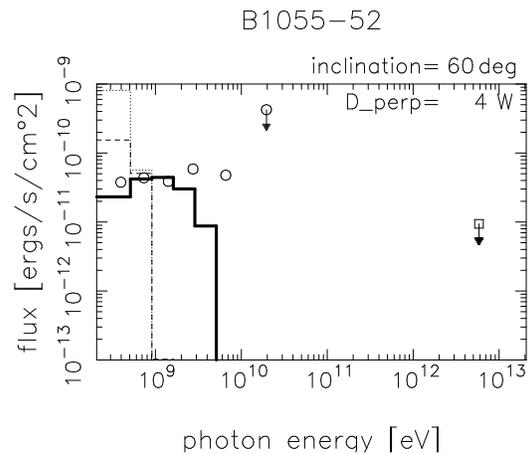} } 
\caption{\label{fig:Spc_1055_60_01_00_j2}
Expected pulsed $\gamma$-ray spectra from PSR~B1055-52
for $\inc= 60^\circ$, $j_{\rm gap}=0.01$, and $j_1=0$.
The four curves correspond to the same cases as presented in \figB.
In each case, the larger flux for 
either the inwardly or the outwardly propagating $\gamma$-rays
is depicted.
If the curve is thick (or thin), 
it means that the outwardly (or inwardly) propagating $\gamma$-ray flux
is greater than the other.
        }
\end{figure}

The emitted spectra are presented in figure~\ref{fig:Spc_1055_60_01_00_j2}.
Because the Lorentz factor decreases as $j_2$ increases,
the spectrum becomes soft as the gap shifts inwards.
For the dotted curve, 
the spectrum would be further softened from 
figure~\ref{fig:Spc_1055_60_01_00_j2}, 
if we considered the unsaturated effect of particle motion.
We can therefore conclude that 
the gap should not be located well inside of the null surface 
for $\inc \approx 60^\circ$, 
so that the predicted GeV spectrum may not be significanly inconsistent 
with observations.

\subsubsection{Current Injection across Inner Boundary}
\label{sec:out_B1055}

\begin{figure} 
\centerline{ \epsfxsize=8.5cm \epsfbox[200 20 500 250]
             {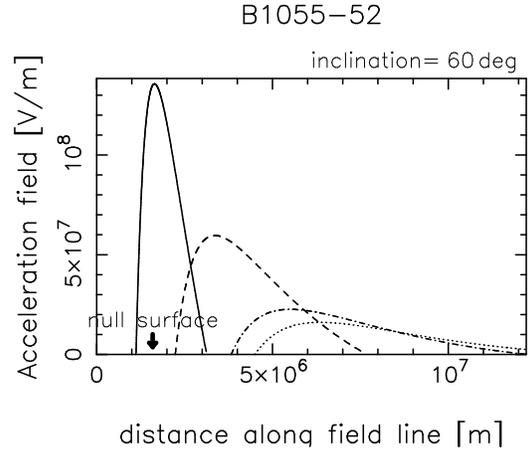} } 
\caption{\label{fig:Ell_1055_60_01_j1_00}
Distribution of $-d\Psi/ds$ for PSR~B1055-52 for $\inc= 60^\circ$. 
The solid, dashed, dash-dotted, and dotted curves correspond to
the cases of ($j_{\rm gap}$,$j_1$,$j_2$)$=$(0.01,0,0),
(0.01,0.25,0), (0.01,0.5,0), and (0.01,0.585,0), respectively.
        }
\end{figure}

Let us consider the case when the gap shifts towards the 
light cylinder, 
as a result of the particle injection across the inner boundary.
In figure~\ref{fig:Ell_1055_60_01_j1_00},
we present $-d\Psi/ds$ for vanishing $j_2$;
the solid, dashed, dash-dotted, and dotted curves
represent the cases of
($j_1$,$j_2$)= (0,0), (0.25,0), (0.5,0), and (0.585,0), respectively.
For all the four cases, we set
$\inc=60^\circ$ and $j_{\rm gap}=0.01$.
The abscissa covers the range from $s=0$ to $s= 1.3 \rlc$.
There is no solution above $j_1>0.585$, 
because $-d\Psi/ds$ distribution
forms a brim at the outer boundary at this critical value.

It follows from the figure that the gap shifts outwards as $j_1$ increases.
For example, the gap center 
is located at $s=s_{\rm cnt} \equiv (s_1+s_2)/2 = 0.225\rlc$ 
for $j_1=0$ (solid curve), 
while it shifts to the position 
$s=0.526\rlc$ and $s=0.854\rlc$ for $j_1=0.25$ (dashed) 
and $j_1=0.5$ (dash-dotted), respectively.
In addition, $W$ increases as the gap shifts outwards.
This is because the diluted X-ray field at the outer part of the
gap increases $\lambda_{\rm p}$ (see eq.~[\ref{eq:closure}]).
The increase of $W$ with increasing $j_1$, however,
does not mean that $-d\Psi/ds$ increases with $j_1$.
This is because the small 
$\vert \rhoGJ \vert \propto \vert B_z \vert \sim r^{-3}$ 
at larger distance from the star 
results in a small $\vert d(-d\Psi/ds)/ds \vert$ 
in the Poisson equation~(\ref{eq:Poisson_0}).

\begin{figure} 
\centerline{ \epsfxsize=8.5cm \epsfbox[200 20 500 250]
             {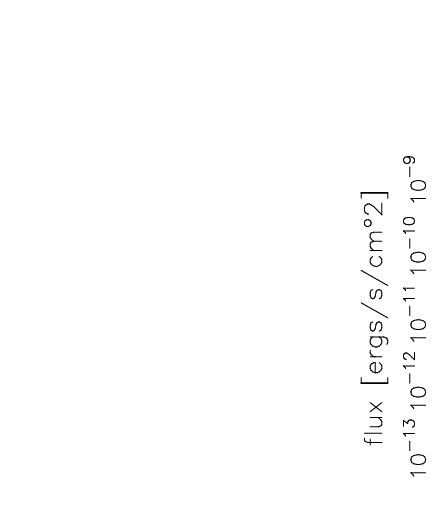} } 
\caption{\label{fig:Spc_1055_60_01_j1_00}
Expected pulsed $\gamma$-ray spectra from PSR~B1055-52
for $\inc= 60^\circ$, $j_{\rm gap}=0.01$, and $j_2=0$.
The solid, dashed, dash-dotted, and dotted curves correspond to
the cases of $j_1= 0, 0.25, 0.5$, and 0.5847, respectively.
In each case, the larger flux for 
either the inwardly or the outwardly propagating $\gamma$-rays
is depicted.
If the curve is thick (or thin), 
it means that the outwardly (or inwardly) propagating $\gamma$-ray flux
is greater than the other.
        }
\end{figure}

The emitted spectra are presented in figure~\ref{fig:Spc_1055_60_01_j1_00}.
The curvature spectrum becomes the hardest for ($j_1$,$j_2$)=(0,0) 
of the four cases,
because $-d\Psi/ds$ becomes large
by virtue of the strong magnetic field at relatively small 
distance from the star.
It follows from the solid and dashed curves in 
figures~\ref{fig:Spc_1055_60_01_00_j2} and \ref{fig:Spc_1055_60_01_j1_00}
that the spectrum does not soften very rapidly when $j_1$ increases
(fig.~\ref{fig:Spc_1055_60_01_j1_00})
compared with the case when $j_2$ increases
(fig.~\ref{fig:Spc_1055_60_01_00_j2}).
This is because $W$ increases with increasing $j_1$
(fig.~\ref{fig:Ell_1055_60_01_j1_00}),
while it significantly decreases with increasing $j_2$
(fig.~\ref{fig:Ell_1055_60_01_00_j2}).

Let us summarize the main points that have been made 
in \S\S~\ref{sec:j1=j2_B1055}--\ref{sec:out_B1055}.
The curvature spectrum becomes the hardest when $j_1=j_2=0$.
In this case, the gap is located close to the null surface,
where $B_z$ vanishes.

\subsection{Dependence on Inclination}
\label{sec:incl_B1055}

\begin{figure} 
\centerline{ \epsfxsize=8.5cm \epsfbox[200 20 500 250]
             {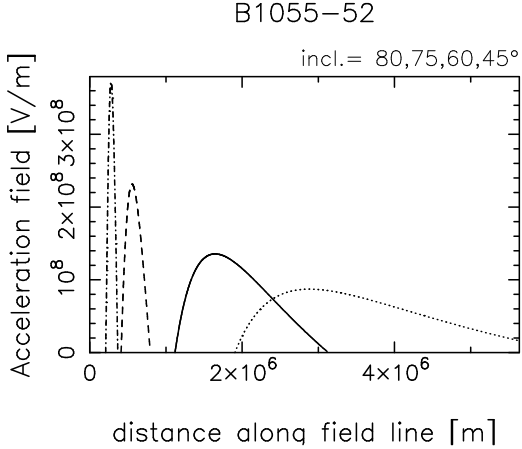} } 
\caption{\label{fig:Ell_1055_4inc}
Distribution of $-d\Psi/ds$ for PSR~B1055-52 for 
$\inc= 80^\circ$ (dash-dotted),
$75^\circ$ (dashed),
$60^\circ$ (solid), and 
$45^\circ$ (dotted). 
The current densities are fixed as 
$j_{\rm gap}=0.01$ and $j_1=j_2=0$;
therefore, the peak of each curve corresponds to the null surface
for each inclination.
        }
\end{figure}

\begin{figure} 
\centerline{ \epsfxsize=8.5cm \epsfbox[200 20 500 250]
             {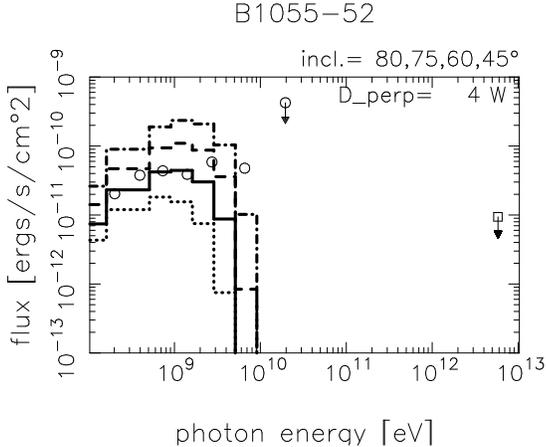} } 
\caption{\label{fig:Spc_1055_4inc}
Expected pulsed $\gamma$-ray spectra from PSR~B1055-52
for $45^\circ$ (dotted),
$\inc= 60^\circ$ (solid), 
$75^\circ$ (dashed), and
$80^\circ$ (dash-dotted).
The current densities are fixed as 
$j_{\rm gap}=0.01$ and $j_1=j_2=0$.
In all the four cases, 
outwardly propagating fluxes are greater than (or dominate) 
the inwardly propagating ones.
        }
\end{figure}

It has been revealed that the $\gamma$-ray spectrum becomes hardest
when the injected current is small (say, when $j_1 \sim j_2 \sim 0$)
and that the EGRET spectrum cannot be explained for $\inc=60^\circ$
(as the solid line in figs.~\ref{fig:Spc_1055_60_01_j1=j2}, 
\ref{fig:Spc_1055_60_01_00_j2}, or \ref{fig:Spc_1055_60_01_j1_00} 
indicates).
Since the $\gamma$-ray energies are predicted to increase with
increasing $\inc$ in Paper~V, we examine in this section
whether the EGRET pulsed
flux around 6~GeV can be explained if consider a larger $\inc$.
In figures~\ref{fig:Ell_1055_4inc} and \ref{fig:Spc_1055_4inc}, 
we present the 
acceleration fields and the expected spectra for 
$\inc=80^\circ$, $75^\circ$, $60^\circ$, and $45^\circ$
as the dash-dotted, dashed, solid, and dotted lines.
The current densities are chosen as 
$j_{\rm gap}=0.01$, $j_1=j_2=0$.
In each case, the outwardly propagating $\gamma$-ray flux is
depicted, because it is greater than the inwardly propagating one.

It follows from figure~\ref{fig:Spc_1055_4inc} that
a large inclination angle ($75^\circ$ or greater) is preferable
to explain the $\gamma$-ray spectrum around 6~GeV.
The reasons are fourfold (see also \S~6.2 in Paper~VII):\\
$\bullet$ \ 
The distance of the null surface from the star
decreases with increasing $\inc$.\\
$\bullet$ \ 
The magnetic field strength, and hence the Goldreich-Julian charge
density, $\Omega B_z/(2\pi c)$, in the gap increases 
as the distance from the star decreases.\\
$\bullet$ \ 
It follows from the Poisson equation~(\ref{eq:Poisson_0}) that
the derivative of $-d\Psi/ds$ (i.e., $-d^2\Psi/ds^2$)
increases with increasing $B_z$
(i.e., with decreasing distance from the star).\\
$\bullet$ \ 
By virtue of this increasing derivative, 
the acceleration field, $-d\Psi/ds$, at the
gap center increases.
We may notice here that $W$ does not decrease very rapidly with increasing $\inc$,
because of the \lq negative' feedback effect due to the $N_\gamma$
factor in equation~(\ref{eq:closure}).
As a result of this increased $-d\Psi/ds$,
the spectrum becomes hard for a large $\inc$.

It is worth noting that the Tev flux is always negligibly small
compared with the observational upper limit.
This is because the infrared flux,
which is deduced from the Rayleigh-Jeans side of the soft surface
blackbody spectrum, 
declines sharply as $\nu^2$ at small frequency, $h\nu \ll kT_{\rm s}$.

\subsection{Dependence on Distance}
\label{sec:distance}

We have assumed that the distance is $0.5$~kpc so that the
observed area of the soft blackbody emission,
$A_{\rm s}=0.78 A_\ast (d/0.5\mbox{kpc})^2$,
may be less than $A_\ast$.
In this section, we consider the case when the distance is
$1.5$~kpc, as indicated by
Taylor, Manchester and Lyne (1993),
and compare the results with $d=0.5$~kpc case.

\begin{figure} 
\centerline{ \epsfxsize=8.5cm \epsfbox[200 20 500 250]
             {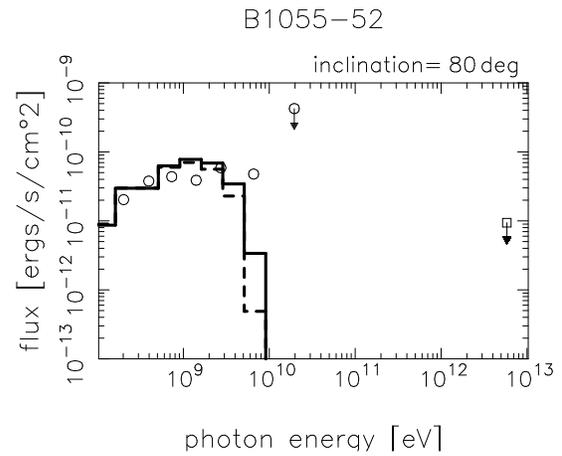} } 
\caption{\label{fig:S1055dist}
Expected pulsed $\gamma$-ray spectra from PSR~B1055-52
for $\inc= 80^\circ$, $j_{\rm gap}=0.01$, and $j_1=j_2=0$.
The solid (or dashed) line represents the $\gamma$-ray spectrum
when the distance is $0.5$~kpc (or $1.5$~kpc).
        }
\end{figure}

In figure~\ref{fig:S1055dist},
we present the resultant spectra for $\inc=80^\circ$,
$j_{\rm gap}=0.01$, and $j_1=j_2=0$.
The solid (or dashed) line represents the $\gamma$-ray spectrum
when the distance is $0.5$~kpc (or $1.5$~kpc).
The normalization is adjusted by the fluxes below $0.5$~GeV;
$D_\perp=2.3W$ (or $7.7W$) is assumed for $d=0.5$~kpc (or $1.5$~kpc).
It is clear from figure~\ref{fig:S1055dist}
that the GeV spectrum becomes hard
if we assume a smaller distance, $d$.

The conclusion that a small distance to this pulsar is preferable,
is derived, 
because the less dense X-ray field for a smaller $d$
results in a larger $\lambda_{\rm p}$ in equation~(\ref{eq:closure}),
and hence in a larger $W$ and the acceleration field, $-d\Psi/ds$.
However, the solution does not vary so much
if the X-ray density increases $(1.5\mbox{kpc}/0.5\mbox{kpc})^2=9$~times.
This is due to the \lq negative feedback effect' caused by
the $N_\gamma^{-1}$ factor in equation~(\ref{eq:closure});
that is, the increase of $W$ is substantially canceled by the increase
of $N_\gamma$, which is proportional to the product of the
curvature-radiation rate of a particle per unit length and $W$.
In other words, the solution exists for a wide range of parameters
and does not change very much if the variation of the X-ray field
density is within one order of magnitude.

\section{Discussion}
\label{sec:discussion}

In summary, we have developed a one-dimensional model for 
an outer-gap accelerator in the magnetosphere of a rotation-powered pulsar.
Solving the Vlasov equations that describe a stationary pair-production
cascade, we revealed that the accelerator shifts 
towards the star surface (or the light cylinder)
as the particle injection rate across the outer (or inner) boundary
approaches the Goldreich-Julian value.
Applying this theory to a middle-aged pulsar, B1055-52,
we find that stationary solutions exist for a wide range of parameters 
such as the injected currents, magnetic inclination, and the distance to 
the pulsar.
Comparing the expected spectrum with ERGET observations,
we conclude that a sub Goldreich-Julian current density,
a large magnetic inclination, and a small distance (500~pc, say)
are preferable for this pulsar.

\subsection{Magnetospheric Current Distribution}
\label{sec:solution_space}

As demonstrated in \S~6.3 in Paper~VII,
the spin-down luminosity of a rotation-powered pulsar becomes
(see \S~6.3 in Paper VII)
\begin{equation}
  \dot{E}_{\rm rot} 
  \sim j_{\rm tot} f_{\rm active}^2
       \frac{\Omega^4 \mu_{\rm m}^2}{c^3},
\end{equation}
where $\mu_{\rm m}$ is the neutron star's magnetic dipole moment;
$f_{\rm active}$ represents the ratio of magnetic fluxes
along which currents are flowing
and those that are open.
For PSR~B1055-52,
$\Omega^4 \mu_{\rm m}^2 / c^3 = 10^{34.64} \mbox{ergs s}^{-1}$.
If all the open field lines are active (i.e., $f_{\rm active}=1$), 
$j_{\rm tot} \sim 1$ is required 
so that the observed spin-down luminosity 
$10^{34.48} \mbox{ergs s}^{-1}$ may be realized.
If a small fraction of open field lines are active
(i.e., $f_{\rm active} \ll 1$),
a super-Goldreich-Julian current (i.e., $j_{\rm tot} \gg 1$) is required.

However, our outer-gap model requires
(and, in fact, previous outer-gap models assume) 
$j_{\rm tot} \ll 1$.
Thus the deficient current must flow along the open field lines
that are not threading the accelerator.

\subsection{Comparison with Previous Works}
\label{sec:cf_previous}

Let us compare the present methods and results with Paper~V. 
In the present paper, $\Ell$, $N_\pm(s)$, and $G_\pm(s,\epsilon_\gamma)$ 
were solved from the Vlasov equations for a non-vacuum gap, 
while in Paper~V only $\Ell$ field was solved from the Poisson equation
for a vacuum gap, with the aid of the 
\lq closure condition'~(\ref{eq:closure}),
where $j_{\rm gap}=j_{\rm tot}$ was assumed.
Since the condition~(\ref{eq:closure}) is automatically satisfied by
the Vlasov equations,
the electrodynamics solved in this paper
(e.g., $W$, $\Ell$) are essentially the same with Paper~V,
provided that the gap is nearly vacuum (i.e., $j_{\rm tot} \ll 1$).
For example, the peak energy of curvature radiation is predicted
to be about $1$~GeV for $\inc=45^\circ$ in Paper~V,
which is comparable with the $\nu F_\nu$ peak 
of the dotted line in figure~\ref{fig:Spc_1055_4inc}.
It also follows from figure~\ref{fig:Spc_1055_4inc} that
the peak energy does not strongly depend on $\inc$;
this conclusion is consistent with what was obtained
for middle-aged pulsars in Paper~V (see table~3).
In the present paper, we can more accurately predict the
$\gamma$-ray spectrum, which is divided into 11 energy bins,
compared with the gray approximation adopted in Paper~V.
For example, the $\gamma$-ray spectrum is predicted to decline 
very sharply above $10$~GeV (fig.~\ref{fig:Spc_1055_4inc})
even for a large $\inc$ ($\sim 75^\circ$, say).

We finally compare the present work with the numerical simulation
performed by Smith et al. (2001),
who examined equilibrium charge distributions in the magnetosphere
of an aligned rotator.
In the first place, they demonstrated that the Goldreich-Julian
charge distribution is unstable and collapses to form a polar dome
containing plasma of one charge and an
equatorial belt containing plasma of the other sign:
$\bf{E}\cdot\bf{B} = 0$ inside both of them.
These are separated by a vaccum gap in which 
$\bf{E}\cdot\bf{B} \ne 0$ holds.
To apply the theory developed in the present paper to the case when
the Goldreich-Julian charge distribution breaks down,
we must modify the Poisson equation~(\ref{eq:Poisson_0})
(or eqs.~[\ref{eq:basic-1}] and [{eq:basic-2}])
so that the electric field caused by the charges in the dome
and the belt may be taken into account.

In the second place, Smith et al. (2001) demonstrated
that the creation of electron-positron pairs in the gap
between the dome and the belt reduces the value of 
$\bf{E}\cdot\bf{B}$ in the gap 
so that it turns off at last.
This is because the charges are attracted to regions of the same sign,
following the closed field lines,
and increasingly filling the magnetosphere.
Instead, as considered in the present paper,
if pairs are created in the open field line region,
we may expect global flows of charged particles
and hence a stationary pair production cascade in the magnetosphere
of a spinning neutron star.
If we extend the work by Smith et al (2001) into oblique rotators
and apply the present method to the gap,
we may construct a stationary magnetospheric model with
pair production cascade under the existence of global currents.
There is room for further investigation on this issue.

\par
\vspace{1pc}\par

One of the authors (K. H.) wishes to express his gratitude to
Drs. Y. Saito and A. K. Harding for valuable advice. 
He also thanks the Astronomical Data Analysis Center of
National Astronomical Observatory, Japan for the use of workstations.

\end{document}